\documentclass[structabstract]{aa}
\usepackage{graphicx} 
\usepackage{txfonts} 

\begin{document}

\title{Mixing in classical novae: a 2-D sensitivity study}

\author{Jordi Casanova \inst{1,2}
   \and Jordi Jos\'e \inst{1,2}
   \and Enrique Garc\'\i a--Berro \inst{3,2}
   \and Alan Calder \inst{4}
   \and Steven N. Shore \inst{5} }

\offprints{J. Jos\'e}

 \institute{Departament de F\'\i sica i Enginyeria Nuclear, EUETIB,
            Universitat Polit\`ecnica de Catalunya, 
            c/Comte d'Urgell 187, 
            E-08036 Barcelona, 
            Spain\
            \and 
            Institut d'Estudis Espacials de Catalunya, 
            c/Gran Capit\`a 2-4, 
            Ed. Nexus-201, 
            E-08034 Barcelona, 
            Spain\
            \and
            Departament de F\'\i sica Aplicada,
            Universitat Polit\`ecnica de Catalunya, 
            c/Esteve Terrades 5, 
            E-08860 Castelldefels, 
            Spain\
            \and
            Department of Physics \& Astronomy, 
            Stony Brook University, 
            Stony Brook, NY 11794-3800\
	    \and
            Dipartimento di Fisica ``Enrico Fermi'',
            Universit\`a di Pisa and INFN, Sezione di Pisa, 
            Largo B. Pontecorvo 3, I-56127 Pisa, Italy
      \email{jordi.jose@upc.edu}}
       
\date{\today}

\abstract{Classical novae  are explosive phenomena that  take place in
          stellar binary  systems.  They are powered  by mass transfer
          from a low-mass, main sequence  star onto a white dwarf. The
          material  piles   up  under  degenerate   conditions  and  a
          thermonuclear  runaway ensues.  The energy  released  by the
          suite of  nuclear processes operating at  the envelope heats
          the material up to peak temperatures of $\sim (1 - 4) \times
          10^8$ K.  During these events,  about $10^{-4} -  10 ^{-5}\,
          M_{\sun}$,  enriched  in  CNO  and  other  intermediate-mass
          elements,  are  ejected  into  the interstellar  medium.  To
          account for the  gross observational properties of classical
          novae  (in  particular,  a  metallicity enhancement  in  the
          ejecta above  solar values), numerical  models assume mixing
          between  the  (solar-like)  material  transferred  from  the
          companion and the outermost  layers (CO- or ONe-rich) of the
          underlying white dwarf.}
         {The  nature of  the mixing  mechanism that  operates  at the
          core-envelope  interface has  puzzled  stellar modelers  for
          about   40  years.   Here   we  investigate   the  role   of
          Kelvin-Helmholtz  instabilities as  a natural  mechanism for
          self-enrichment   of  the   accreted   envelope  with   core
          material.}
         {The feasibility of this mechanism is studied by means of the
          multidimensional code FLASH. Here,  we present a series of 9
          numerical  simulations perfomed in  two dimensions  aimed at
          testing the  possible influence of  the initial perturbation
          (duration,  strength, location,  and  size), the  resolution
          adopted, or the  size  of the  computational  domain on  the
          results.}
         {We  show that  results do  not depend  substantially  on the
          specific  choice  of  these parameters,  demonstrating  that
          Kelvin-Helmholtz   instabilities  can   naturally   lead  to
          self-enrichment of the accreted envelope with core material,
          at levels that agree with observations.}
         {}


\titlerunning{Mixing in novae: a 2-D sensitivity study} 
\authorrunning{J. Casanova et al.} 

\maketitle

\section{Introduction}

Classical  novae  are cataclysmic  stellar  events. Their  thermonuclear
origin, theorized  by Schatzmann (1949, 1951) and  Cameron (1959) ---
see also Gurevitch \& Lebedinsky (1957) and references therein --- has
been  established through  multiwavelength observations  and numerical
simulations pioneered by Sparks  (1969), who performed  the first
1-D, hydrodynamic nova simulation.   These efforts helped to establish
a  basic  picture,  usually  referred  to as  the  {\it  thermonuclear
  runaway} model  (TNR), which has  been successful in  reproducing the
gross observational properties of  novae, namely the peak luminosities
achieved, the abundance pattern, and the overall duration of the event;
see Starrfield  et al. (2008),  Jos\'e \& Shore (2008),  Jos\'e \&
Hernanz (2007) for recent reviews.

Many details of the dynamics of nova explosions remain to be explored.
In particular, there are  many observed cases of nonspherical ejecta,
inferred from  line profiles during  the early stages of  the outburst
and from  imaging of  the resolved ejecta,  including 
multiple      shells,      emission      knots,      and      chemical
inhomogeneities. Although the broad  phenomenology of the outburst can
be captured by 1-D calculations, it is increasingly clear
that the  full description requires  a multidimensional hydrodynamical
simulation  of   such  outbursts.   To  match  the   energetics,  peak
luminosities,  and the  abundance pattern,  models of  these explosions
require  mixing of  the material  accreted from  the  low-mass stellar
companion with  the outer  layers of the  underlying white  dwarf.  In
fact, because of the moderate  temperatures achieved during the TNR, a
very limited production of elements beyond those from the CNO-cycle is
expected (Starrfield et al. 1998, 2009; Jos\'e \& Hernanz 1998; Kovetz
\&  Prialnik 1997;  Yaron  et  al. 2005),  and  the specific  chemical
abundances derived from observations (with a suite of elements ranging
from H to  Ca) cannot be explained by thermonuclear processing
of  solar-like  material.    Mixing  at  the  core-envelope  interface
represents a likely mechanism.

The details of  the mixing episodes by which  the envelope is enriched
in metals have challenged theoreticians  for nearly 40  years. Several
mechanisms  have  been  proposed, including  diffusion-induced  mixing
(Prialnik \& Kovetz  1984; Kovetz \& Prialnik 1985;  Iben et al. 1991,
1992;  Fujimoto  \& Iben  1992),  shear  mixing  at the  disk-envelope
interface (Durisen  1977; Kippenhahn  \& Thomas 1978;  MacDonald 1983;
Livio \& Truran  1987; Kutter \& Sparks 1987;  Sparks \& Kutter 1987),
convective  overshoot-induced flame  propagation  (Woosley 1986),  and
mixing by gravity wave breaking  on the white dwarf surface (Rosner et
al.  2001; Alexakis  et  al.  2004).  The  multidimensional nature  of
mixing has  been addressed by Glasner  \& Livne (1995)  and Glasner et
al. (1997, 2005, 2007) with 2-D simulations of CO-novae performed with
VULCAN, an arbitrarily Lagrangian  Eulerian (ALE) hydrocode capable of
handling both explicit and  implicit steps. They report an effective
mixing  triggered  by  Kelvin-Helmholtz  instabilities  that  produced
metallicity    enhancements    to    levels    in    agreement    with
observations. Similar studies (using the same initial model as Glasner
et al. 1997)  were conducted by Kercek et al. (1998,  1999) in 2-D and
3-D,  respectively. Their  results,  computed with  the Eulerian  code
PROMETHEUS,  displayed  mild TNRs  with  lower  peak temperatures  and
velocities than Glasner et  al. (1997) and insufficient mixing.  While
Glasner  et al. (1997)  argue that  substantial mixing  can naturally
occur  close to  peak  temperature, when  the  envelope becomes  fully
convective and drives  a powerful TNR, Kercek et  al. (1998) conclude
instead that mixing must take  place much earlier:  if it occurs
around peak temperature, it leads  to mild explosions or to events that
do not resemble a nova.

The  differences  between these  studies  have been  carefully analyzed  by
Glasner  et al. (2005),  who conclude  that the  early stages  of the
explosion, before TNR ignition when the evolution is quasi-static, are
extremely  sensitive to  the adopted  outer boundary  conditions. They
show that  Lagrangian simulations, in which the  envelope is allowed
to expand  and mass  is conserved, lead  to consistent  explosions. In
contrast, in  Eulerian schemes with a ``free  outflow'' outer boundary
condition, the choice  adopted in  Kercek et  al.  (1998),  the
outburst can  be artificially quenched. The scenario  was revisited by
Casanova et  al. (2010), who  show that simulations with  an Eulerian
scheme  --- the  FLASH  code ---  and  a proper  choice  of the  outer
boundary conditions can produce deep-mixing of the solar-like accreted
envelopes with core material.  The puzzling results reported by Kercek
et al.   (1998) stress the need  for a systematic  evaluation of the
effect that different choices  of model parameters (e.g. the intensity
and location  of the initial temperature  perturbation, resolution, or
size of  the computational domain) may  have on the  results.  To this
end,  we performed  a series  of 9  numerical simulations  in 2-D
aimed at  testing the  influence of these  parameters on the  level of
metal enhancement of the envelope. Here we report the results of these
simulations. 

Our paper  is organized  as follows. In Sect. 2  we explain  our input
physics and initial  conditions. Then Sect. 3 is  devoted to studying
the mixing at  the core-envelope interface for our  fiducial model. In
Sect.  4 the  effects  of the  size  of the  initial perturbation  are
analyzed, while in  Sect. 5 we discuss the effects of  the size of the
computational domain.  In Sect.   6 we quantify  the influence  of the
grid resolution.  Finally, in Sect.   7 we discuss the significance of
our results and draw our conclusions.

\section{Input physics and initial conditions}

\begin{figure}
\centering
\includegraphics[width=\columnwidth]{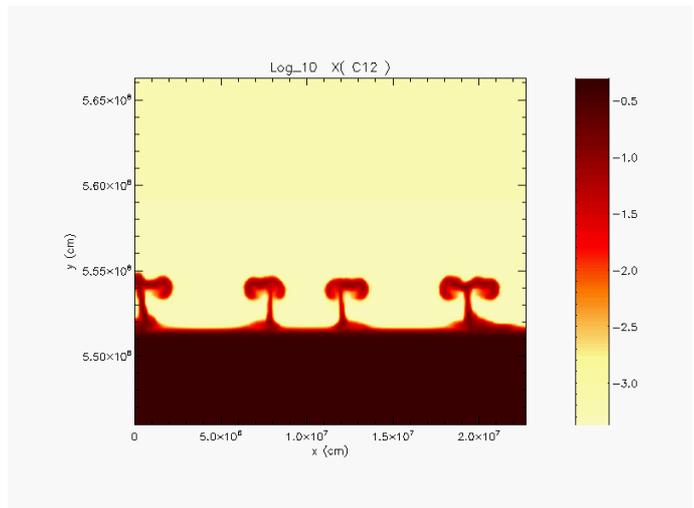}
\caption{Snapshot  of  the development  of  early instabilities, which 
         later spawn Kelvin-Helmholtz instabilities, shown in terms of
         the $^{12}$C  mass fraction (in logarithmic  scale) for model
         A,  158  s  from  the   start  of  the  simulation  when  the
         core-envelope  interface temperature  is  $T_{\rm base}  \sim
         1.36 \times 10^8$ K.}
\label{fig:KH}
\end{figure}

The   simulations  reported   here  were   performed  with   FLASH,  a
parallelized explicit Eulerian code,  based on the piecewise parabolic
interpolation  of physical quantities  for solving  the hydrodynamical
equations, and  with adaptive mesh  refinement (see Fryxell et al. 2000). 
As in Casanova  et al.
(2010), we   used the same initial model as  Glasner et al. (1997)
and  Kercek et  al. (1998):  a  $1 \,  M_{\sun}$ CO  white dwarf  that
accretes solar  composition matter ($Z=0.02$)  at a rate of  $5 \times
10^{-9}\, M_{\sun}$ yr$^{-1}$. The model was evolved spherically (1-D)
and mapped onto a 2-D cartesian grid, when the temperature at the base
of the envelope  reached $\approx 10^8$ K. It  initially comprised 112
radial layers --- including the outermost  part of the CO core --- and
512 horizontal layers. The mass  of the accreted envelope was about $2
\times  10^{-5}\,  M_{\sun}$.  Nuclear  energy  generation is  handled
through  a  network  of  13  species  ($^1$H,  $^{4}$He,  $^{12,13}$C,
$^{13,14,15}$N,  $^{14,15,16,17}$O,  and  $^{17,18}$F), and  connected
through 18 nuclear reactions.  We adopted the conductive and radiative
opacities from Timmes (2000) and  an equation of state based on Timmes
\& Swesty  (2000). Periodic boundary  conditions were imposed on both
sides   of  the   computational  domain   with   vertical  hydrostatic
equilibrium  with an outflow constraint at the  top and a reflecting
constraint at the  bottom on the velocity (see
Zingale et al. 2002).  A summary of the main characteristics of the 9
models computed  in this work  is given in  Table 1, where $H$  is the
distance from the perturbation to the initial core-envelope interface,
$R_x$ and $R_y$\footnote{The different values adopted for
$R_x$  and $R_y$  in  models  F and  G  are used  to  account for  the
assumption  of  a   rectangular  (rather  than  square)  computational
domain.},  $\delta T$,  and  $\delta  t$ are  the  size, strength,  and
duration of  the temperature  perturbation, and $Z$  the mass-averaged
metallicity of the envelope at the end of the calculations.

\begin{table*}
\caption{Models computed.}
\label{table:1}      
\centering                                  
\begin{tabular}{c c c c c c c c c c}          
\hline
\hline        
Model & H  & $R_x \times R_y$ & $\delta T$ & $\delta t$ & Resolution & Computational & $t_{\rm KH}$ & $t_Y$ & $Z$ \\
      &(km)& (km)   &           & (s)       & (km)     & Domain  (km)    & (s)  & (s)  &  \\
\hline                                 
  A & 0  & 1 $\times$ 1     &  5\%   &$10^{-10}$ & 1.56 $\times$ 1.56 &  800 $\times$  800 & 155 & 496 & 0.224 \\
  B & 0  & 1 $\times$ 1     &  5\%   &$10$       & 1.56 $\times$ 1.56 &  800 $\times$  800 &  28 & 347 & 0.212 \\
  C & 0  & 1 $\times$ 1     &  0.5\% &$10^{-10}$ & 1.56 $\times$ 1.56 &  800 $\times$  800 & 155 & 493 & 0.209 \\
  D & 5  & 1 $\times$ 1     &  5\%   &$10^{-10}$ & 1.56 $\times$ 1.56 &  800 $\times$  800 & 154 & 496 & 0.235 \\
  E & 5  & 5 $\times$ 5     &  5\%   &$10^{-10}$ & 1.56 $\times$ 1.56 &  800 $\times$  800 & 156 & 486 & 0.209 \\
  F & 0  & 2 $\times$ 1     &  5\%   &$10^{-10}$ & 1.56 $\times$ 1.56 & 1600 $\times$  800 & 151 & 493 & 0.206 \\
  G & 0  & 1 $\times$ 1.25  &  5\%   &$10^{-10}$ & 1.56 $\times$ 1.56 &  800 $\times$ 1000 & 156 & 526 & 0.291 \\
  H & 0  & 1 $\times$ 1     &  5\%   &$10^{-10}$ & 1 $\times$ 1       &  800 $\times$  800 & 162 & 584 & 0.201 \\
  I & 0  & 1 $\times$ 1     &  5\%   &$10^{-10}$ & 0.39 $\times$ 0.39 &  800 $\times$  800 & 268 & 893 & 0.205 \\
\hline
\hline      
\end{tabular}
\end{table*}

\begin{figure*}
\centering
\includegraphics[width=0.45\textwidth]{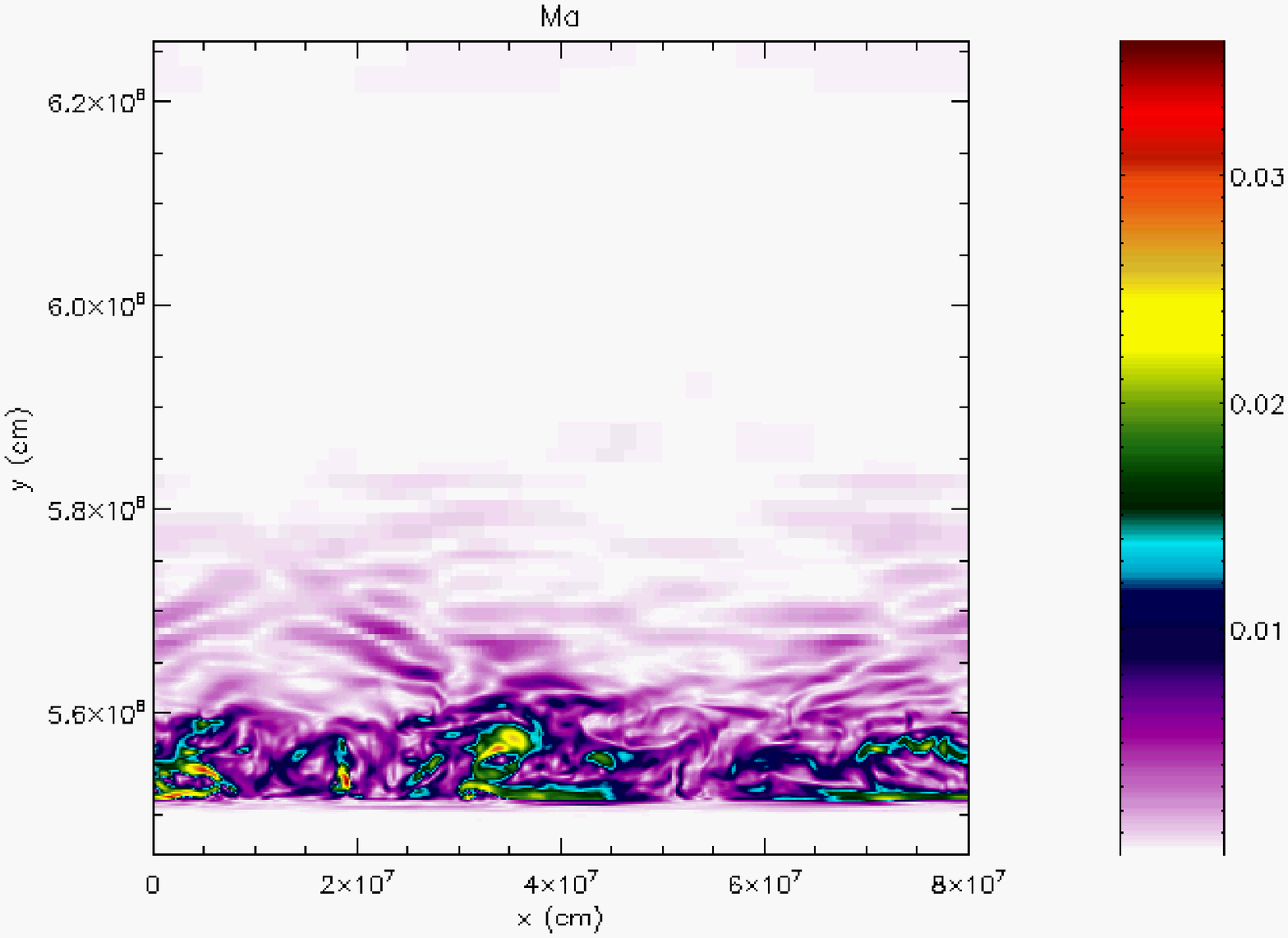}
\includegraphics[width=0.45\textwidth]{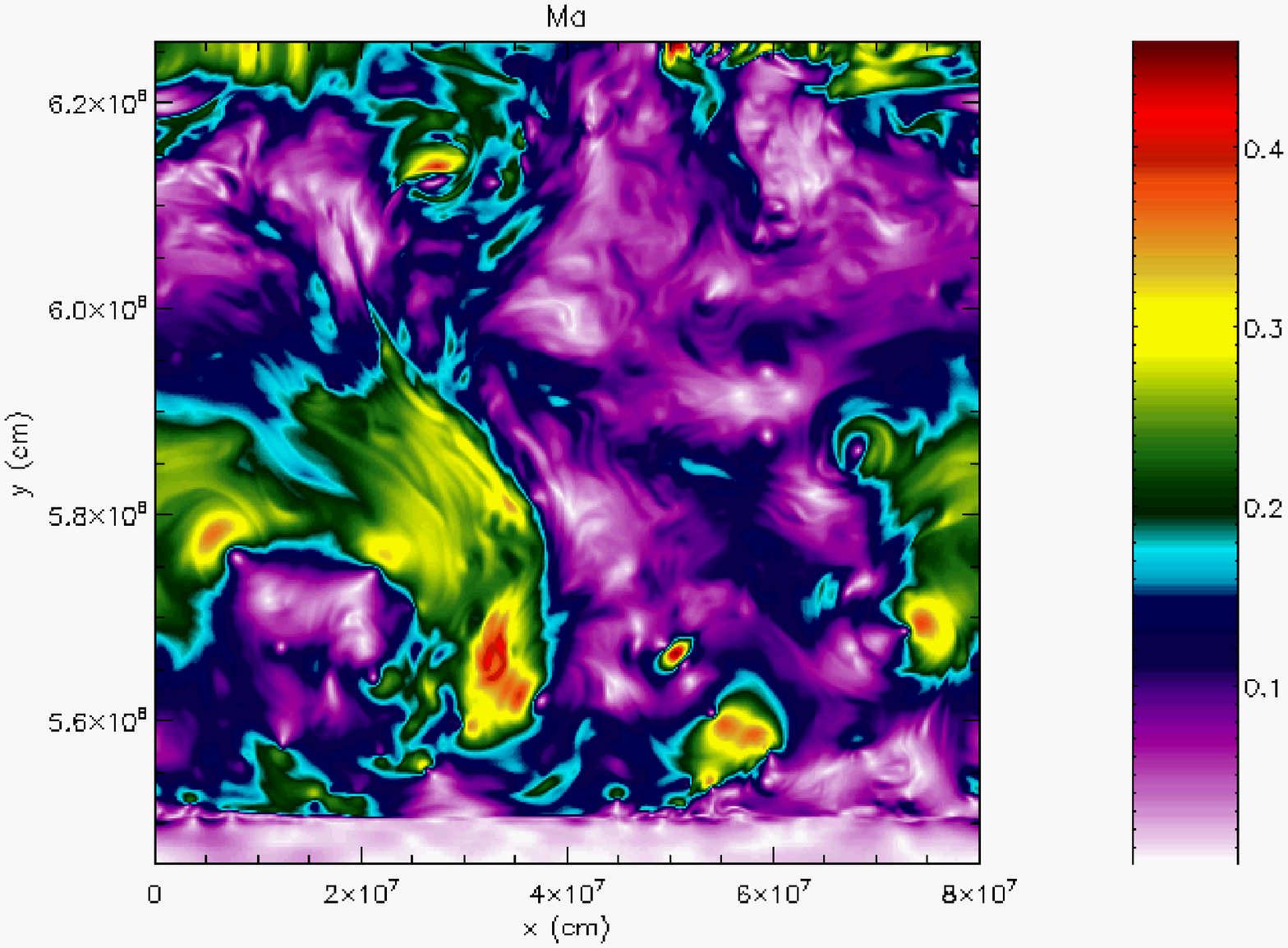}
\caption{Mach  number  at two  different  moments  of the  simulation,
         $t=230$ s (left panel) and 496 s (right panel), for model A.}
\end{figure*}

\section{2-D simulations of mixing at the core-envelope interface}

In this Section, we describe  the basic features of our fiducial model
A,  as  a  framework for  further  discussion  of  the effect  of  the
parameter choices on our results.  A movie, showing the development of
Kelvin-Helmholtz instabilities,  in terms of the  $^{12}$C content, up
to the  time when  the convective front  hits the  upper computational
boundary, ModelA-2D.wmv, is               available                 at
http://www.fen.upc.edu/users/jjose/Downloads.html.                 The
simulation was performed for the  conditions of model A, as summarized
in Table 1.

For all sequences reported in this work, the relaxation of the initial
model to  guarantee hydrostatic  equilibrium, together with  the small
amount  of numerical viscosity  --- in  contrast with  the simulations
performed  by   Glasner  et  al.   (1997)  ---   requires  an  initial
perturbation close to the core-envelope interface to trigger the onset
of instabilities early in  the calculations.  The initial perturbation
is  applied  to  the  temperature  using  four  parameters:  strength,
location,  size  and  duration.   Model  A  assumes  a  {\it  top-hat}
temperature perturbation wherever $((x-x_0)/R_{x})^2
+  ((y-y_0)/R_{y})^2  \leq  1$,  where  $x$  and  $y$  are  the  space
coordinates  measured from  the  center of  the perturbation,  ($x_0$,
$y_0$), and  $R_x$ and  $R_y$ indicate its  spatial extent.   We fixed
$x_0$ =  $5 \times  10^7$ cm  in all sequences.   The strength  of the
perturbation is 5\% in temperature in all cases but one (see table 1).
It is 1 km wide,  applied only during the initial timestep 
(that is, the temperature is fixed only during $10^{-10}$ s), 
and imposed on the  core-envelope interface ($y_0$ = 5.51 $\times$
$10^8$ cm).  The  resolution adopted in model A  is $1.56 \times 1.56$
km, and the size of the computational domain is $800 \times 800$ km.

\begin{figure*}
\centering
\includegraphics[angle=-90,width=0.45\textwidth]{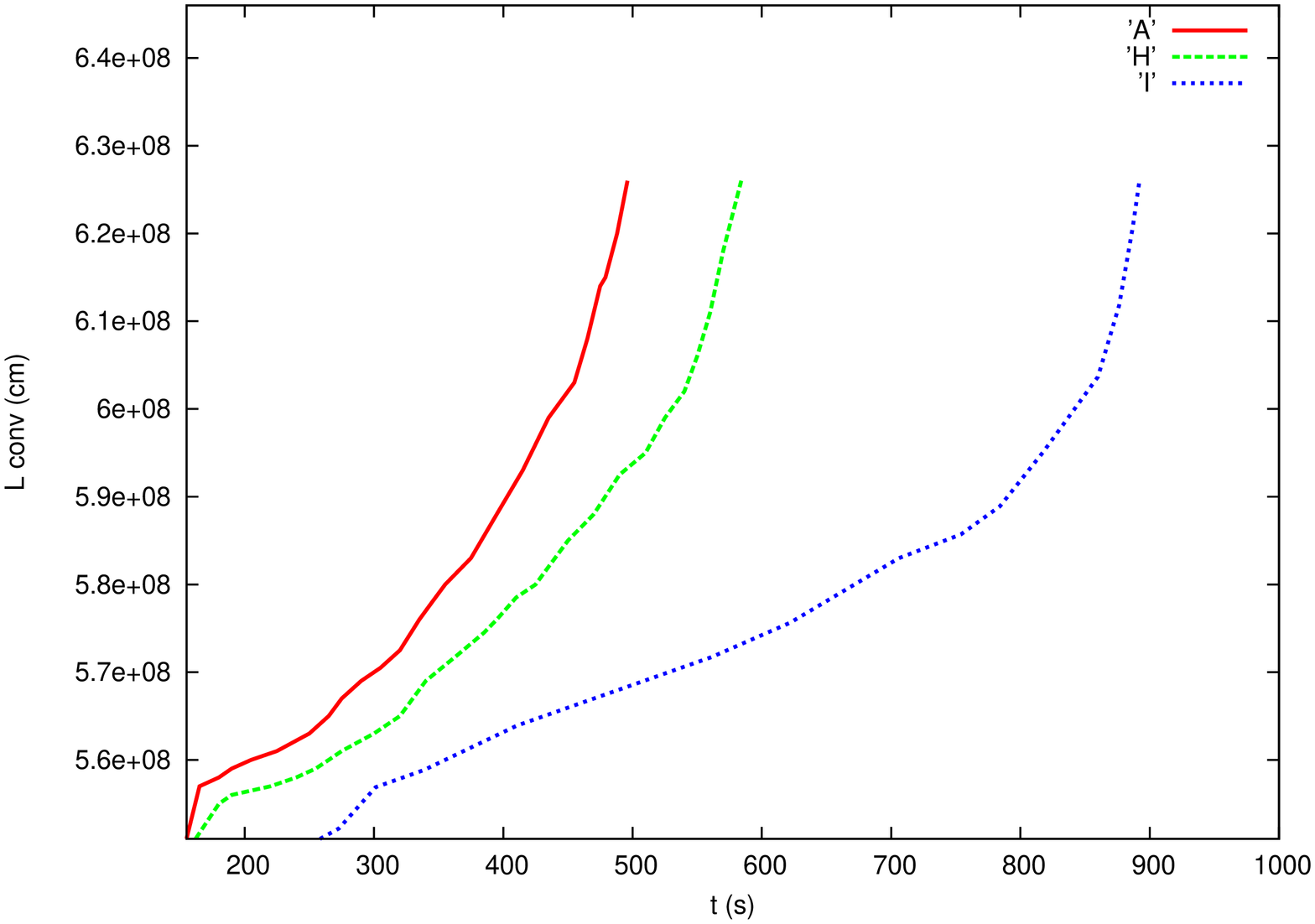}
\includegraphics[angle=-90,width=0.45\textwidth]{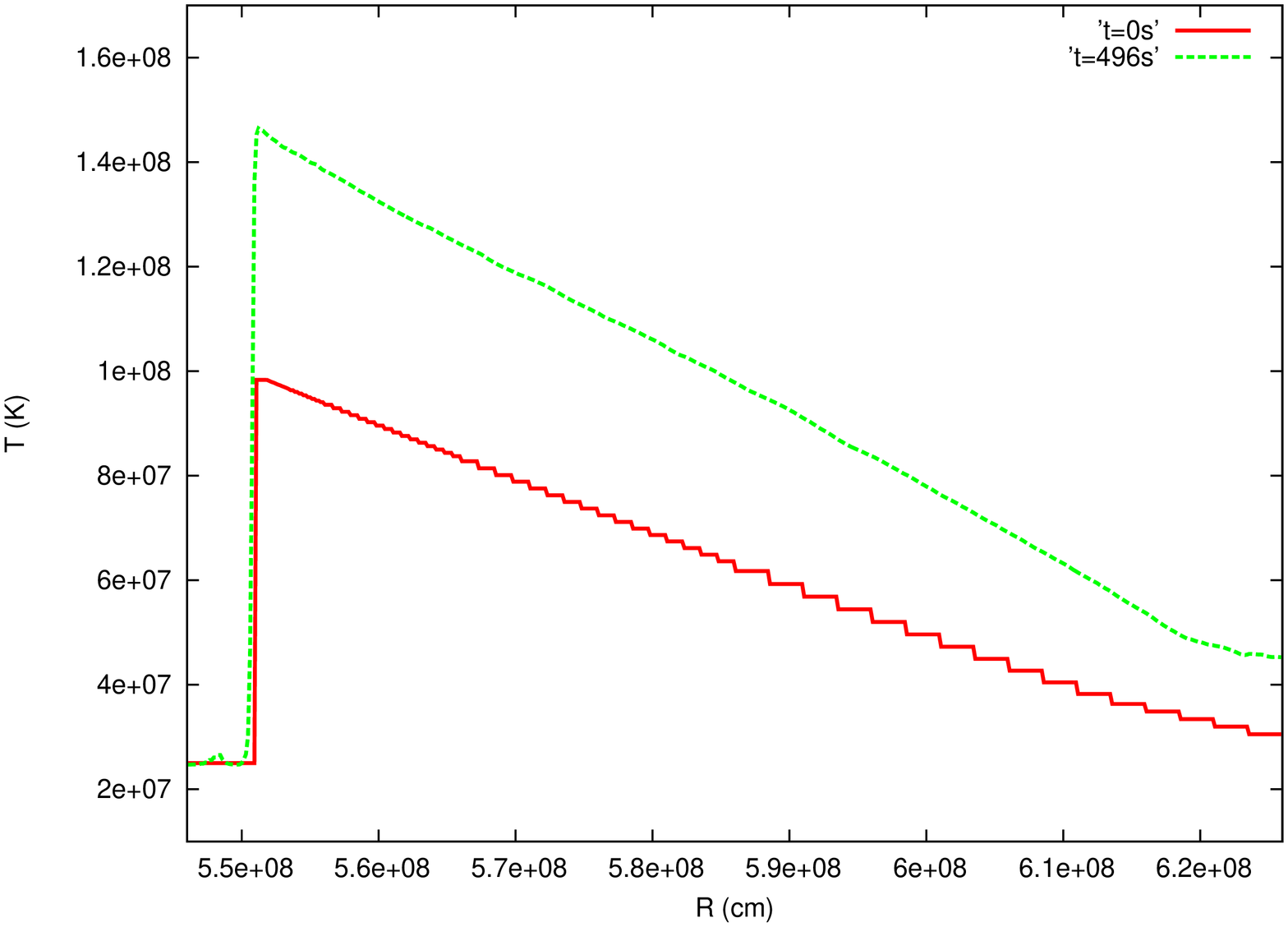}
\caption{Left panel: propagation of the convective front as a function
         of time,  for models  A, H, and  I. Right  panel: temperature
         profile versus radius at  two different times, $t=0$ s (solid
         line;  $T_{\rm  base}=9.84  \times  10^7$ K)  and  $t=496$  s
         (dashed line;  $T_{\rm base}=1.64 \times 10^8$  K), for model
         A.}
\label{fig:advance}
\end{figure*}

The  initial  perturbation  drives  a  shear flow  that  triggers  the
formation of  instabilities (Fig. 1), about  150 s after  the start of
the simulation.  As soon as material  from the core is  mixed into the
envelope,  small convective cells  develop. At  this early  stage, the
fluid has a  large Reynolds number, with a  characteristic eddy length
of 50  km, fluid  velocities ranging  between $v$ =  $10^5 -  10^6$ cm
s$^{-1}$, and  a dynamic  viscosity\footnote{The dynamic viscosity
evaluates the resistance to flow of a fluid under an applied force. 
More precisely, it is defined as the tangential force per unit area required 
to move one horizontal plane with respect to the other at unit velocity when 
maintaining a unit distance apart by the fluid.} of $2  \times 10^4$ P
The fluid
velocity $v$  remains below the speed  of sound $c_{\rm  s}$ (that is,
the  Mach number  Ma$=v/c_{\rm  s}$  is always  less  than unity,  see
Fig.  2), hence,  the  fluid  displays a  deflagration  rather than  a
detonation ---  see Williams  (1985), for a  thorough analysis  of the
differences   between   flame   propagation   under   detonation   and
deflagration  conditions. At  $t  = 235$  s,  The characteristic  eddy
turnover time is  $l_{\rm conv}/v_{\rm conv} \sim$ 10  s.  The merging
of the small convective cells into large eddies, characteristic of 2-D
simulations, with  a size  comparable to the  height of  the envelope,
reinforces    the   injection   of    CO-rich   material    into   the
envelope. Convection becomes more turbulent.  At this stage ($t = 450$
s), the nuclear energy  generation rate exceeds 10$^{15}$ erg g$^{-1}$
s$^{-1}$, while  the characteristic convective  timescale decreases to
$\sim$  5 s.   The convection  front propagates  progressively upwards
(Fig.  3,  left  panel),  with  a  velocity of  $\sim$  10  km/s,  and
eventually reaches  the top of our computational  domain. The envelope
base reaches a peak temperature of  $1.64 \times 10^8$ K. At this time
($t = 496$  s), when matter starts to cross the  outer boundary of the
computational domain, we stop the calculations because of the Eulerian
nature of the FLASH code.  At this final stage, the mean mass-averaged
metallicity in the envelope reaches $Z \sim 0.21$. It is worth noting,
however,  that  the convective  eddies  are  still pumping  metal-rich
matter through the core-envelope  interface.  Hence, it is likely that
the final metallicity  in the envelope will be  larger. The simulation
shows that the induced Kelvin-Helmholtz vortices can naturally lead to
self-enrichment of the accreted  envelope with core material to levels
that agree with observations and  that the expansion and progress of
the runaway  is almost spherically symmetric for  nova conditions even
for a point-like TNR ignition.

\section{Effect of the initial perturbation}

To quantify the influence of  the initial perturbation on our results,
we have  performed a  series of  2-D hydrodynamic tests  for a  set of
different durations,  strengths (intensities), locations  and sizes of
the  perturbation.   For  simplicity,  a {\it  top-hat}  perturbation,
centered at $x_0  = 5 \times 10^7$ cm, has been  adopted in all models
reported in this work.

The effect of the duration of the perturbation was checked by means of
a test  case (model B), identical  to model A but  with a perturbation
lasting for 10 s.  As  shown in Table 1, the characteristic timescales
for model B, such as the  time required for the first instabilities to
show up, $T_{\rm  KH}$, or the time needed by  the convective front to
hit the  outer boundary, $t_Y$, become  shorter. The role  played by a
temperature  perturbation can  be understood  in terms  of  the energy
injected  into   the  envelope:  the   longer  the  duration   of  the
perturbation, the  larger the energy  injected, and thus,  the shorter
the  characteristic timescales of  the TNR.   This has  little effect,
however, on the overall  metallicity enhancement in the envelope since
a  final CNO  mass fraction  of  $\sim 0.212$  was found  in model  B,
whereas $\sim 0.224$ resulted in model A.

Both models  A and  B assumed temperature  perturbations of  $\delta T
\sim 5$\%  during the initial  timestep ($\sim 10^{-10}$ s).   To test
the possible  influence of  the strength of  the perturbation,  a test
case with $\delta T \sim 0.5$\%  (model C) has also been computed.  As
shown in Table 1  and Fig. 4, the time evolution of  models A and C is
very similar, and hence, similar  final mean CNO mass fractions at the
end of the simulations were found (with $Z = 0.209$ in model C).

The effect of the location of the perturbation along the vertical axis
has  also  been  studied:   whereas  model  A  assumed  a  temperature
perturbation of  $\sim$ 5\%, applied  at the innermost  envelope shell
($y_0 = 5.51 \times 10^8$ cm),  in model D, a similar perturbation was
placed $\sim$  5 km  above the core-envelope  interface ($y_0  = 5.515
\times  10^8$  cm).  Both  models  exhibit  a  very  similar  temporal
evolution, with almost identical times for the appearance of the first
instabilities  and   for  the  time   required  to  reach   the  outer
boundary. Similar  envelope mean CNO mass fractions  (0.224 and 0.235,
respectively) were also found.

Finally, the influence  of the size of the  perturbation has also been
analyzed.   Whereas model D  was evolved  with an  initial temperature
perturbation of size  $R_x$ = 1 km  and $R_y$ = 1 km,  model E assumed
$R_x$ = 5 km and $R_y$  = 5 km. As before, very similar characteristic
timescales (see Table 1) and  final mean CNO mass fractions (0.235 and
0.209, respectively) were found.

To summarize,  the specific choice  of the parameters that  define the
initial   temperature  perturbation   has  a   negligible   effect  on
metallicity enhancement of the envelope.

\section{Effect of the size of the computational domain}

The choice of the computational domain represents a compromise between
computational  time  requirements  and  numerical  accuracy.   Several
considerations constrain its minimum size.  On one hand, the merger of
large convective eddies often found in 2-D simulations may be severely
affected  by the  adoption of  a small  computational domain.   On the
other hand,  nova outbursts eventually result in  mass ejection.  With
an  Eulerian code  such as  FLASH,  it is  not possible  to track  the
material that  flows off the grid,  and hence, it is  important to use
domains  that are  as large  as  possible along  the radial  direction
(while    being    sufficiently     wide    along    the    horizontal
axis). Unfortunately,  when the initial  1-D model is mapped  into the
2-D grid, and relaxed  to guarantee hydrostatic equilibrium, densities
quickly underflow values for large heights (Zingale et al. 2002).

The  specific size adopted  for most  of the  models computed  in this
work, i.e.  $800  \times 800$ km, is a bit smaller  than those used in
Glasner  et  al.   (1997)  ---  $0.1  \pi^{rad}$,  in  spherical-polar
coordinates --- and in Kercek et al. (1998) --- $1800 \times 1100$ km,
in a cartesian, plane-parallel  geometry.  In this section, we analyze
possible  dependences  of the  results  on  the  adopted size  of  the
computational  domain. To  this end,  two additional  simulations were
performed.  In the  first one (model F), a  wider computational domain
has been adopted  (i.e., $1600 \times 800$ km).   In the second (model
G), aimed at testing the  influence of the vertical (radial) length, a
domain of  $800 \times  1000$ km  has been used  (where the  choice of
$1000$ km results from  numerical restrictions that limit the vertical
extent of our computational domain).

As shown in Table 1, the  horizontal width (model F) has no noticeable
effect  on the  timescales of  the  simulations, either  for the  time
required  for the onset  of the  first instabilities  or for  the time
required for  the convective front  to reach the outer  boundary.  The
mass-averaged CNO  abundance in the  envelope reached $\sim$  0.206 at
the end  of this  simulation, close  to the value  found for  model A.
These results  confirm that  800 km is  an appropriate choice  for the
width of  the computational domain,  stressing that above  a threshold
value the  course of the TNR  is insensitive to the  adopted width, in
agreement  with   the  sensitivity  study  performed   by  Glasner  et
al. (2007).

\begin{figure*}
\centering
\includegraphics[angle=-90,width=0.45\textwidth]{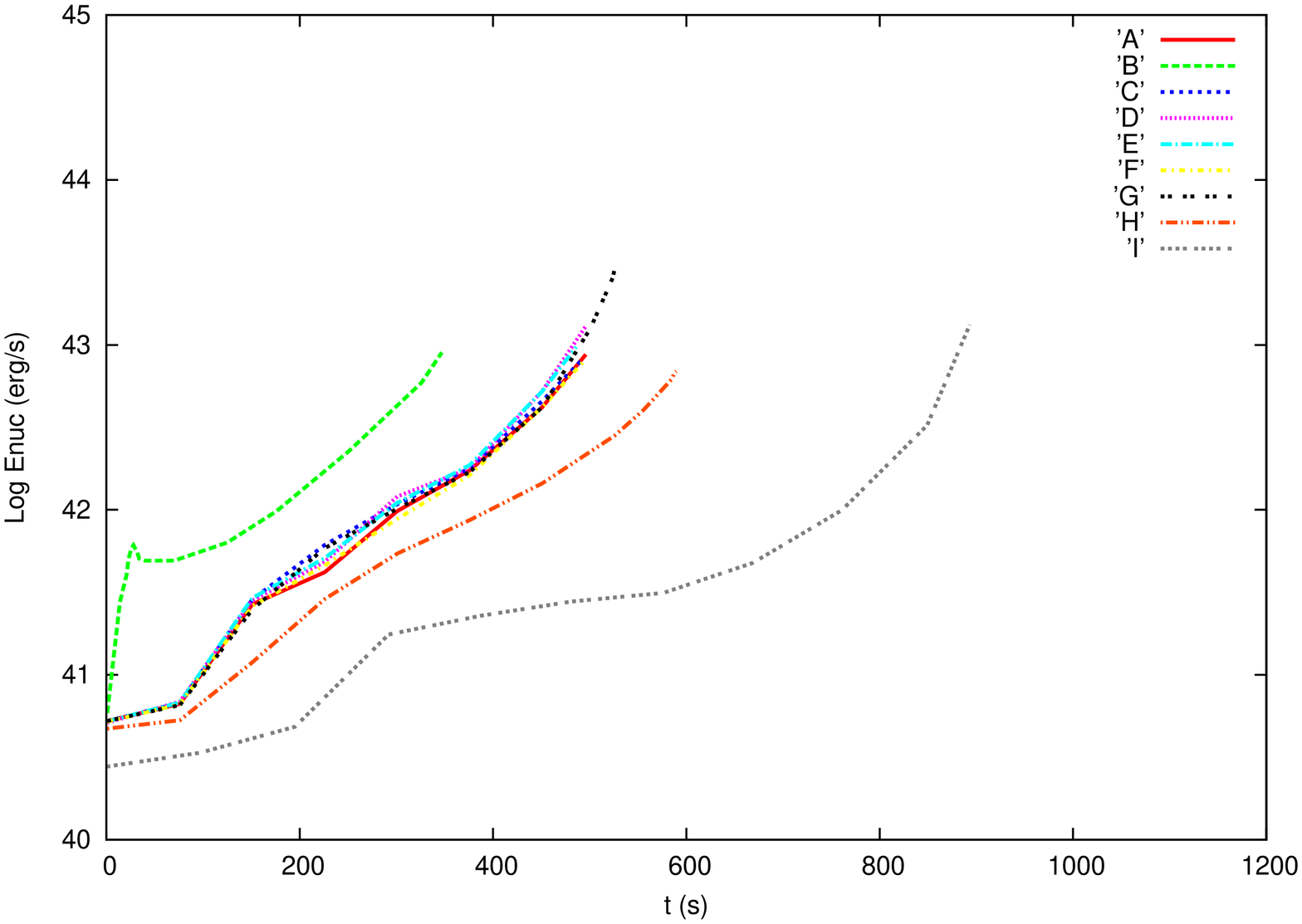}
\includegraphics[angle=-90,width=0.45\textwidth]{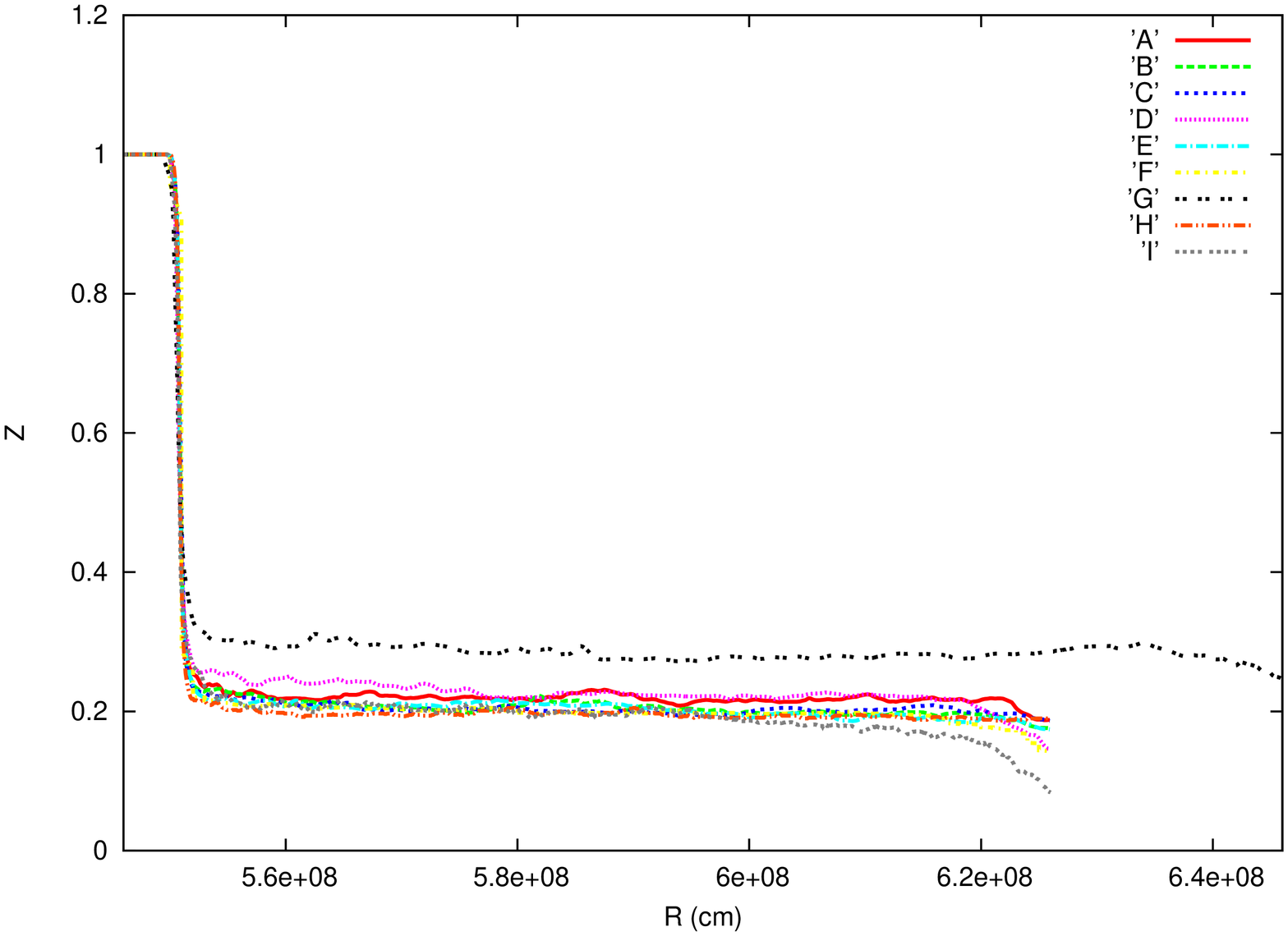}
\caption{Left panel:  time evolution of the  nuclear energy generation
         rate  (in erg.s$^{-1}$)  for the  9 models  computed  in this
         work. Right panel: final CNO mass fraction versus radius.}
\label{fig:enuc_comparison}
\end{figure*}

The specific  length adopted along  the vertical direction  (see model
G), while unimportant for the  time of appearance of the instabilities
(around  155 s  after the  start of  the simulation,  as in  model A),
affects the time required to  reach the outer boundary, located 200 km
above the value  adopted for model A.  Moreover,  the larger extension
of  the computational  domain  along the  radial (vertical)  direction
allows  the  convective  eddies  to pump  additional  metal-rich  core
material into the envelope  compared with all the simulations reported
previously in this paper.  Indeed, the mean, mass-averaged metallicity
in model G achieves the largest value of all the simulations reported,
$\sim 0.291$.  This result suggests that the likely mean mass-averaged
metallicity  driven  by Kelvin-Helmholtz  instabilities  should be  $Z
\approx  0.3$.   In  summary,  we   conclude  that  the  size  of  the
computational  domain, above  a  certain threshold  value, has  little
influence on  the physical quantities  that are more  directly related
with the mixing process at the core-envelope interface.

\section{Effect of the grid resolution}

\begin{figure*}
\centering
\includegraphics[width=0.45\textwidth]{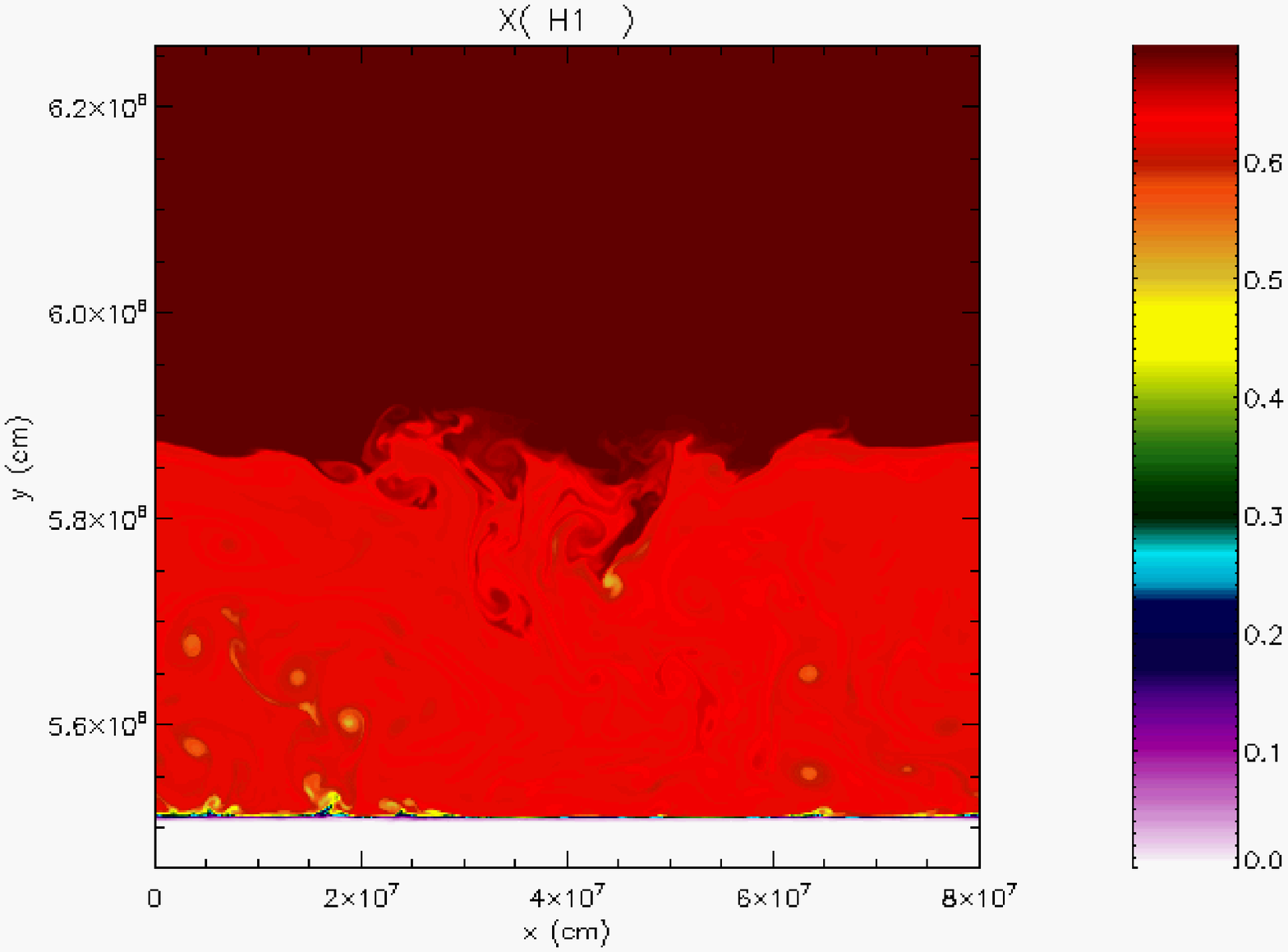}
\includegraphics[width=0.45\textwidth]{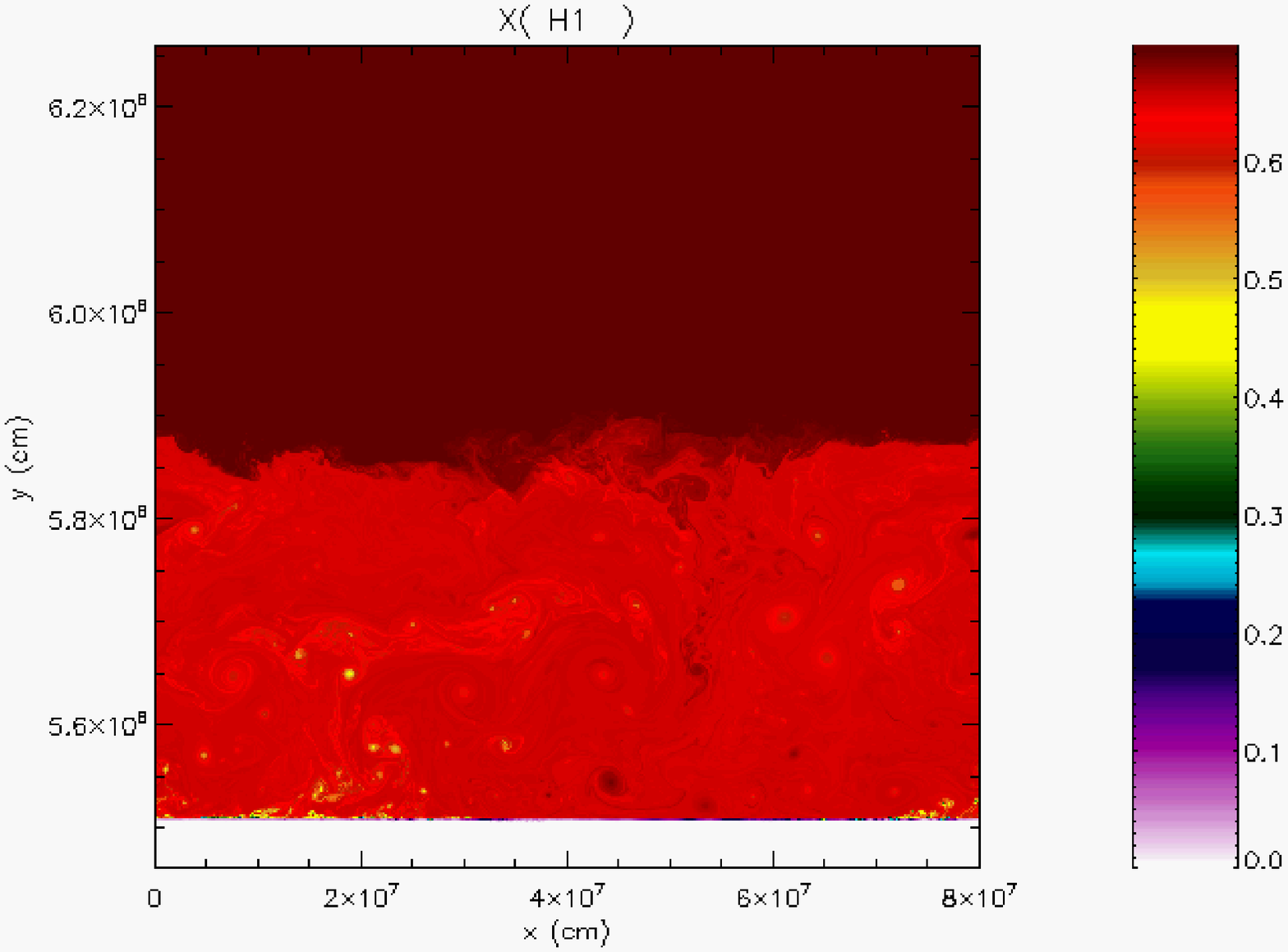}
\includegraphics[width=0.45\textwidth]{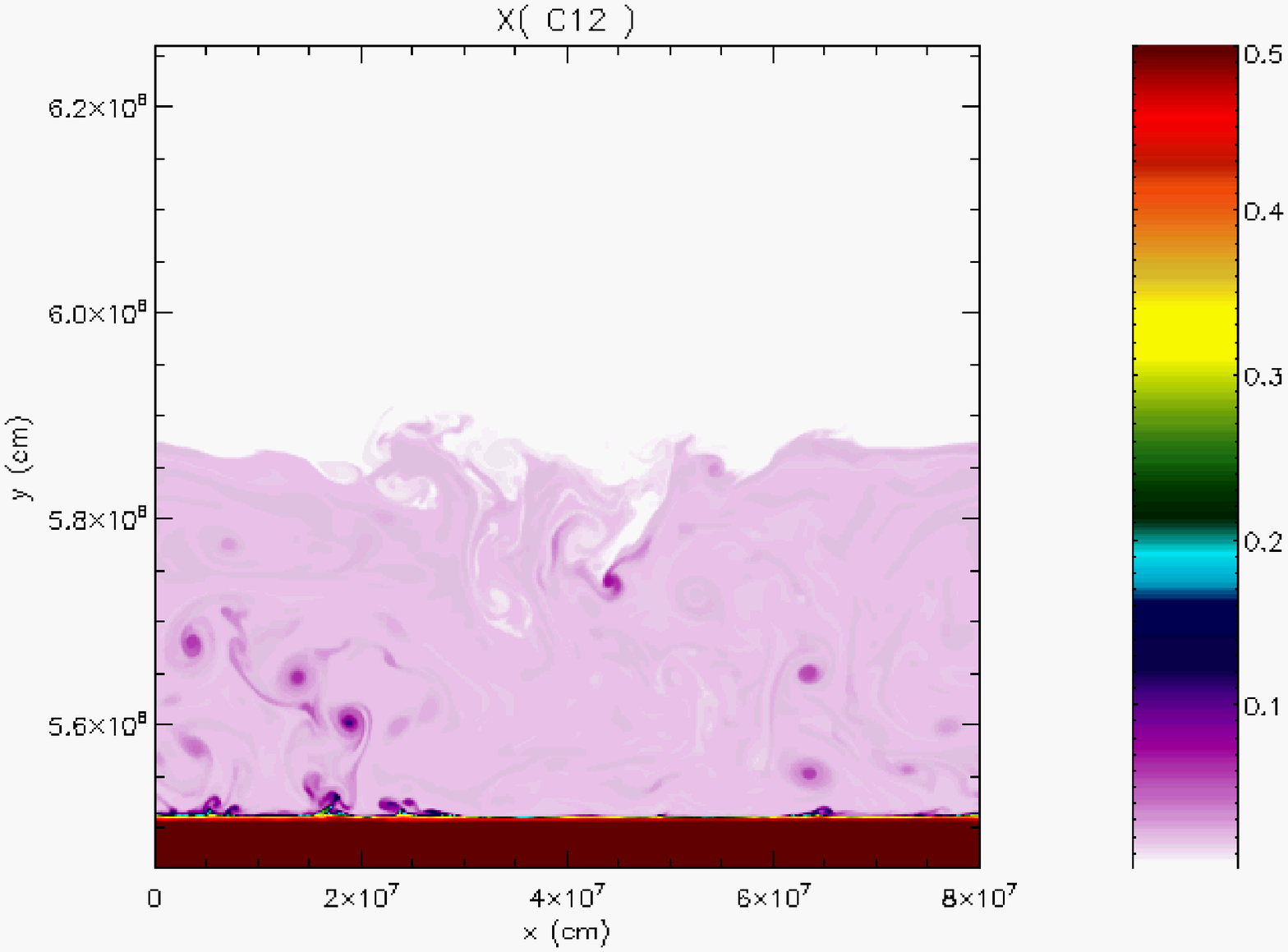}
\includegraphics[width=0.45\textwidth]{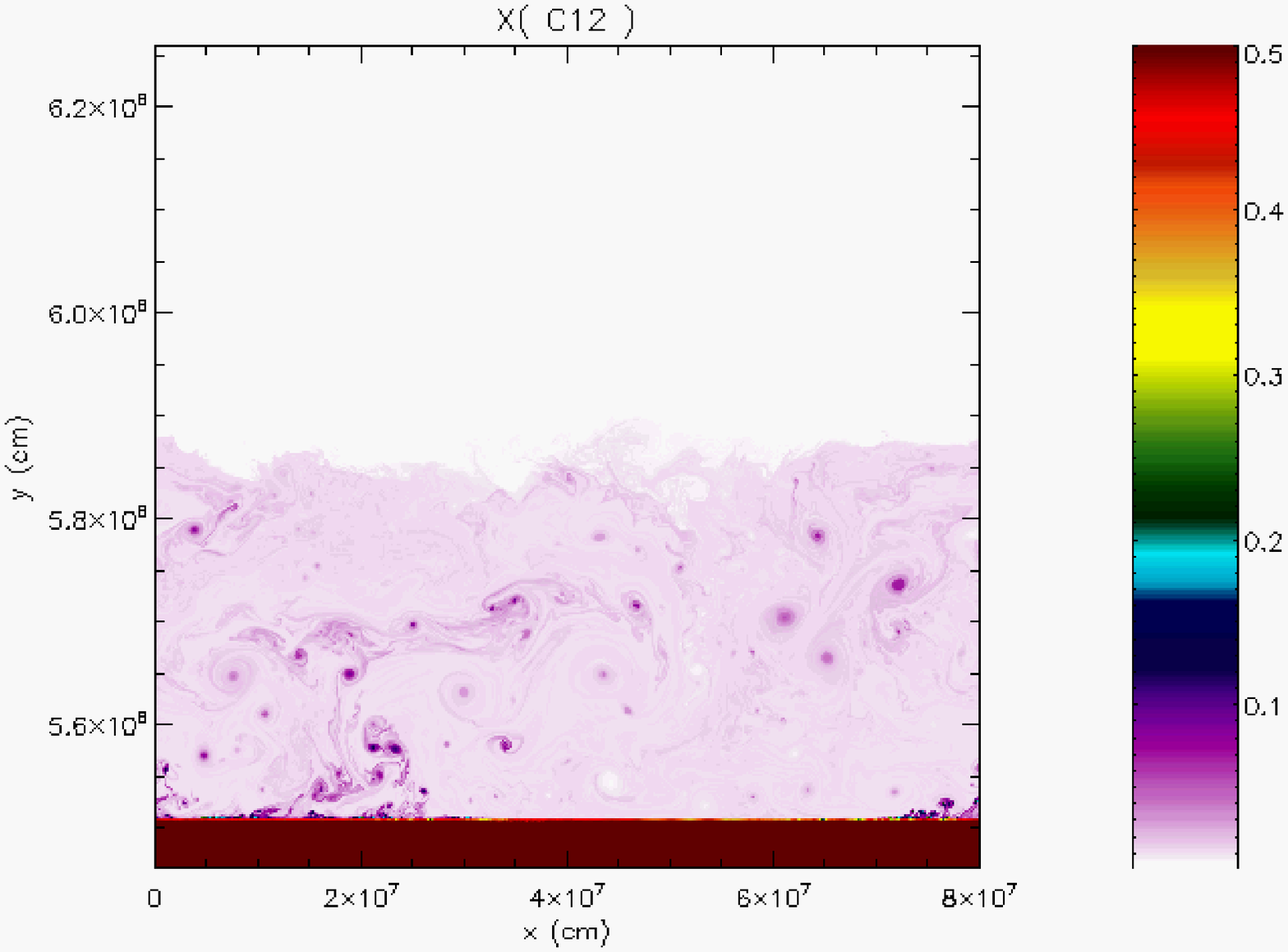}
\includegraphics[width=0.45\textwidth]{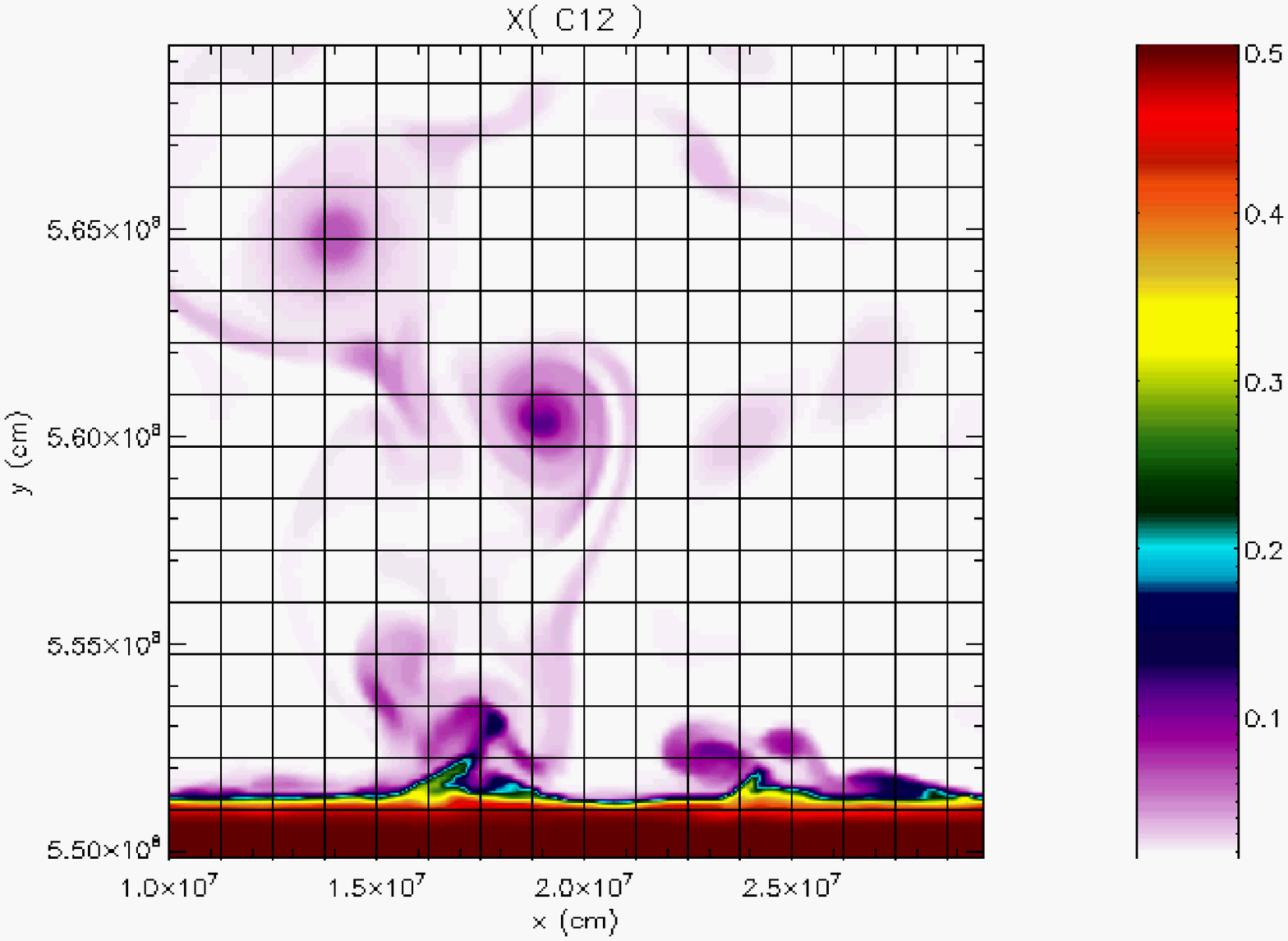}
\includegraphics[width=0.45\textwidth]{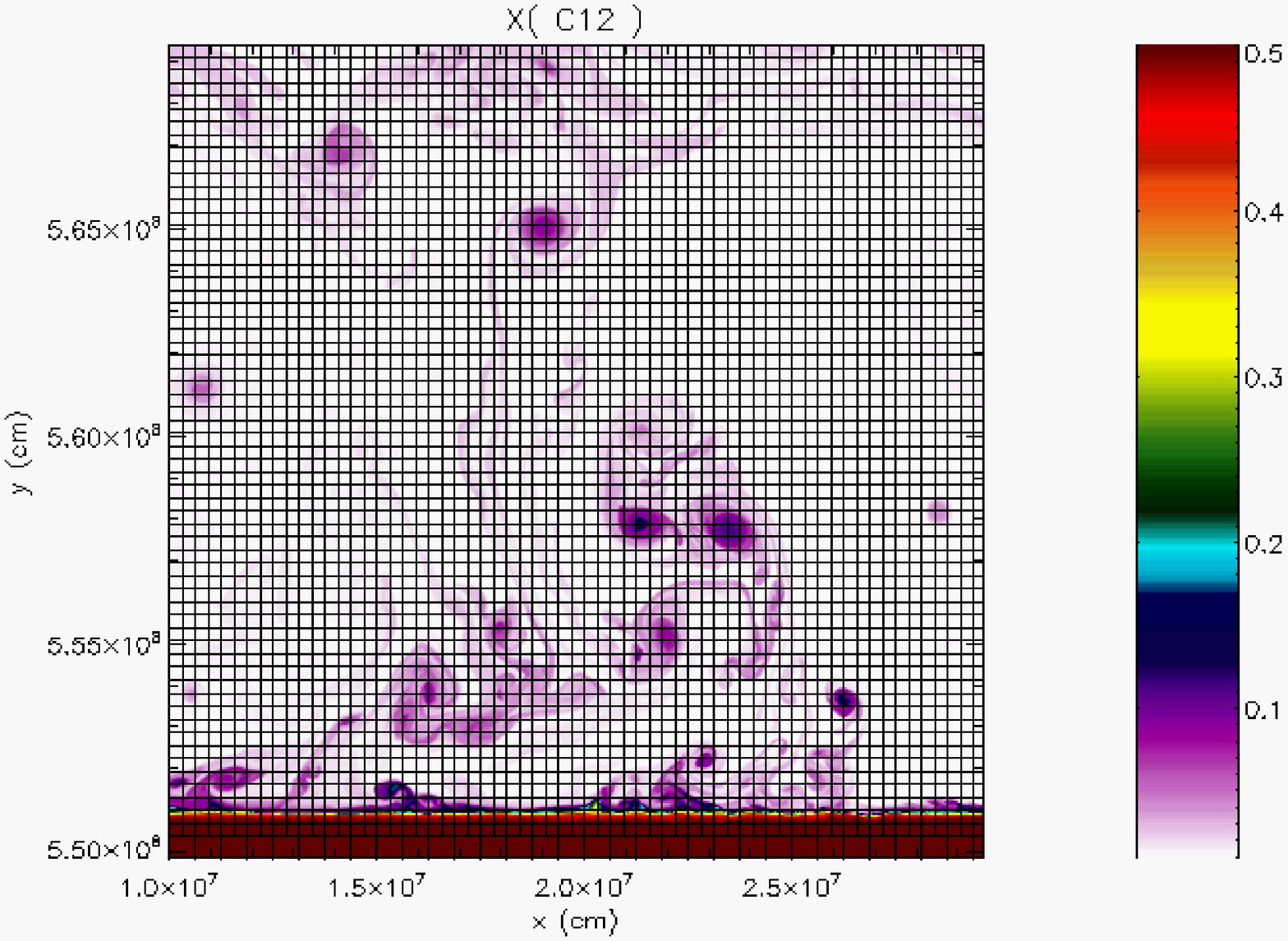}
\caption{Snapshots of the $^{1}$H  (upper panels) and $^{12}$C (middle
         panels)  mass fractions  at $t  \sim$  395 s  (model A;  left
         panels), and 688 s (model I; right panels). Lower panels: the
         number  of blocks administered,  at this  stage, is  3184 for
         model A, and  43800 for model I.  In  both simulations, FLASH
         divides each block in 8 cells. Structures such as vortexs are
         better resolved in the finer resolution model I.}
\label{fig:C12}
\end{figure*}

All simulations discussed so far  (e.g., models A to G) were performed
with a  resolution of $1.56  \times 1.56$ km,  a value similar  to the
minimum resolutions adopted in Glasner  et al. (2007) which is roughly
$\sim 1.4  \times 1.4$ km, and in  Kercek et al. (1998),  $1 \times 2$
km.  To  quantitatively assess the possible effect  of the resolution,
two additional  test cases were  computed with exactly the  same input
parameters as in model A but with two different resolutions: $1 \times
1$  km (model  H) and  $0.39  \times 0.39$  km (model  I)\footnote{For
comparison, whereas  a maximum number of 5300  blocks are administered
in model  A, the number of blocks increases  up to 83000 in  model I. The
total CPU time spent in  both simulations, using 256 processors of the
MareNostrum supercomputer, has been 3 and 110 khr, respectively.}.

As shown  in Table 1, the  increase in resolution produces  a delay in
the  time required  for the  first instabilities  to  develop, $t_{\rm
KH}$. This seems to be a numerical artifact.  In models with a coarser
resolution,  the larger size  of the  blocks artificially  generates a
larger numerical diffusion compared to models with a finer resolution
(a  similar  resolution dependence  is
clearly seen as well in the Kercek et al. 1998 simulations).
Actually, the ratio of differences in the initial build up
times (i.e., (model  I-model A)/(model H-model A)) scales
approximately as the zone size dimensions to the power of two.
This is a purely numerical perturbation that forces the development of
instabilities.   To test  this hypothesis,  we computed  an additional
test case (not included in Table  1), identical to model A but without
any initial perturbation.   The onset of the instabilities  in such an
extremely low numerical diffusion regime is substantially delayed. The
simulations  reported by  Glasner et  al. (1997)  also show  the early
appearance  of instabilities  in  a model  with substantial  numerical
noise: {\it within a very short time (about 10 s), the numerical noise
(round-off) seeds  an intense convective flow in  the envelope without
any  artificial  perturbations}. 

A  similar  behavior is  also  found for  the  time  required for  the
convective  front to  reach the  outer  boundary, $t_Y$,  and for  the
history of the nuclear energy  generation rate (Fig. 4).  As expected,
filamentary structures and convective cells are better resolved in the
finer  resolution model I,  compared to  those computed  with somewhat
coarser grids (models  A and H; see Fig.  5).  These minor differences
do not,  however, show significant  variations in the final,  mean CNO
abundances achieved in  the envelope: while $Z\sim 0.224$  in model A,
models H and  I yield 0.201 and 0.205,  by mass, respectively. Similar
agreement  is found  in  the  peak temperatures  achieved  and in  the
overall nuclear energy generation rates (Fig. 4).

Thus, the adopted resolution has  not a critical effect for the mixing
models presented  in this  work. The variation  in the final  mean CNO
abundance in the envelope, under  the range of resolutions adopted, is
only about 12\% (when comparing results for models A, H, and I), 
a quite reasonable value.
 
\section{Discussion and Conclusions}

In  this  paper  we  have  reported  results  for  a  series  of  nine
2-D  numerical simulations  that test  the influence  of the
initial  perturbation (duration,  strength, location,  and  size), the
resolution of  the grid, and the  size of the  computational domain on
the results.  We have shown that mixing at the core-envelope interface
proceeds almost  independently of the specific choice  of such initial
parameters, above threshold values.

The  study confirms  that  the metallicity  enhancement inferred  from
observations  of the  ejecta of  classical novae  can be  explained by
Kelvin-Helmholtz   instabilities,  powered   by   an  effective   {\it
mesoscopic} shearing  resulting from the initial  buoyancy. Fresh core
material is  efficiently transported from the outermost  layers of the
white dwarf  core and mixed  with the approximately  solar composition
material of the  accreted envelope.  As soon as  $^{12}$C and $^{16}$O
are dredged up, convection sets  in and small convective cells appear,
accompanied by  an increased nuclear energy generation  rate. The size
of these  convective cells increases in time.  Eventually, these cells
merge  into large  convective eddies  with  a size  comparable to  the
envelope   height.    The  range   of   mean  mass-averaged   envelope
metallicities  obtained  in  our  simulations  at the  time  when  the
convective front hits  the outer boundary, $0.21 -  0.29$, matches the
values obtained for classical novae hosting CO white dwarfs.

It is, however,  worth noting that the convective  pattern is actually
produced by the adopted geometry (e.g., 2-D), forcing the fluid motion
to  behave very differently  than 3-D  convection (Shore
2007; Meakin \& Arnett 2007).  Nevertheless, the levels of metallicity
enhancement found in our 2-D simulations will likely remain unaffected
by the  limitations imposed by  the 2-D geometry (D.   Arnett, private
communication). Fully 3-D simulations aimed at testing this hypothesis
are currently underway.

\begin{acknowledgements}  
We thank the referee for a careful reading of the manuscript and
for constructive comments.
The software used in this work was in part developed by the
DOE-supported ASC/Alliances Center for Astrophysical Thermonuclear
Flashes at the University of Chicago.
This  work has  been partially  supported  by the  Spanish MEC  grants
AYA2010-15685 and AYA2008-04211-C02-01, by  the E.U.  FEDER funds, and
by  the ESF EUROCORES  Program EuroGENESIS.   We also  acknowledge the
Barcelona Supercomputing  Center for a generous allocation  of time at
the MareNostrum supercomputer.
\end{acknowledgements}


\begin{thebibliography}{}

\bibitem[]{Ale04} Alexakis, A., et al., 2004, ApJ, 602, 931
\bibitem[]{Cam59} Cameron, A.G.W., 1959, ApJ, 130, 916
\bibitem[]{Cas10} Casanova, J., Jos\'e, J., Garc\'\i a--Berro, E.,
                  Calder, A., \& S. N. Shore, 2010, A\&A, 513, L5
\bibitem[]{Dur77} Durisen, R. H., 1977, ApJ, 213, 145
\bibitem[]{Fry00} Fryxell, B., et al., 2000, ApJS, 131, 273 
\bibitem[]{FI92} Fujimoto, M.-Y., \& Iben, I., Jr., 1992, ApJ, 399, 646
\bibitem[]{GL95}  Glasner, S.A., \& Livne, E., 1995, ApJ, 445, L149
\bibitem[]{GLT97} Glasner, S.A., Livne, E., \& Truran, J.W., 1997, ApJ, 475, 
      754 
\bibitem[]{GLT05} ---. 2005, ApJ, 625, 347
\bibitem[]{GLT07} ---. 2007, ApJ, 665, 1321
\bibitem[]{GL57} Gurevitch, L.Z. \& Lebedinsky, A.I., 1957, in
                 {\it Non-stable stars}, ed. G.H. Herbig 
		 (Cambridge: Cambridge Univ. Press), 77
\bibitem[]{IFM91} Iben, I., Jr., Fujimoto, M.-Y., \& MacDonald, J., 
    1991, ApJ, 375, L27
\bibitem[]{IFM92} ---. 1992, ApJ, 388, 521
\bibitem[]{Jos98} Jos\'e, J., \& Hernanz, M., 1998, ApJ 494, 680 
\bibitem[]{JH07} ---. 2007, J. Phys. G: Nucl. Part. Phys., 34, R431
\bibitem[]{Jos08} Jos\'e, J., \& Shore, S., 2008, in {\it Classical Novae}, 
 ed. M.F. Bode \& A. Evans (Cambridge: Cambridge Univ. Press), 121
\bibitem[]{KHT98} Kercek, A., Hillebrandt, W., \& Truran, J.W., 1998, A\&A, 
        337, 379
\bibitem[]{KHT99} ---. 1999, A\&A, 345, 831
\bibitem[]{KT78} Kippenhahn, R., \& Thomas, H.-C., 1978, A\&A, 63, 265
\bibitem[]{KP85} Kovetz, A., \& Prialnik, D., 1985, ApJ, 291, 812
\bibitem[]{KP97}  ---. 1997, ApJ, 477, 356
\bibitem[]{KS87}  Kutter, G.S., \& Sparks, W.M., 1987, ApJ, 321, 386
\bibitem[]{LT87} Livio, M., \& Truran, J. W., 1987, ApJ, 318, 316
\bibitem[]{Mac83} MacDonald, J., 1983, ApJ, 273, 289
\bibitem[]{MA07} Meakin, C. A., \& Arnett, D., 2007, ApJ, 667, 448
\bibitem[]{PK84} Prialnik, D., \& Kovetz, A., 1984, ApJ, 281, 367
\bibitem[]{Ros01} Rosner, R., Alexakis, A., Young, Y., Truran, J.W., \& 
     Hillebrandt, W., 2001, ApJ, 562, L177 
\bibitem[]{Sch49} Schatzman, E., 1949, Ann. d'Ap., 12, 281
\bibitem[]{Sch51} Schatzman, E., 1951, Ann. d'Ap., 14, 294
\bibitem[]{Sho07} Shore, S.N., 2007, {\it Astrophysical Hydrodynamics}, 
    (Wiley: Darmstadt) 
\bibitem[]{Spa69} Sparks, W.M., 1969, ApJ, 156, 569
\bibitem[]{SK87}  Sparks, W.M., \& Kutter, G.S., 1987, ApJ, 321, 394
\bibitem[]{Sta08} Starrfield, S., Iliadis, C., \& Hix, W.R., 2008, in 
 {\it Classical Novae}, ed. M.F. Bode \& A. Evans (Cambridge: Cambridge 
 Univ. Press), 77
\bibitem[]{Sta98} Starrfield, S., Truran, J.W., Wiescher, M., \& Sparks, W., 
      1998, MNRAS, 296, 502
\bibitem[]{Sta09} Starrfield, S., Iliadis, C., Hix, W.R., Timmes, F.X., \& 
      Sparks, W.M., 2009, ApJ, 692, 1532
\bibitem[]{Tim00} Timmes, F. X., 2000, ApJ, 528, 913
\bibitem[]{TS00}  Timmes, F.X., \& Swesty, F.D., 2000, ApJS, 126, 501  
\bibitem[]{Wil85} William, F.A., 1985, {\it Combustion Theory}, 
            2nd. Ed. (Perseus Books: Reading)
\bibitem[]{Woo86} Woosley, S.E., 1986, in {\it Nucleosynthesis and 
      Chemical Evolution}, eds.: B. Hauck, A. Maeder, and G. Meynet 
       (Geneva Observatory: Sauverny), 1
\bibitem[]{Yar05} Yaron, O., Prialnik, D., Shara, M.M., \& Kovetz, A., 2005, 
        ApJ, 623, 398
\bibitem[]{Zin02} Zingale, M., et al., 2002, ApJS, 143, 539
\end{thebibliography}
\end{document}